\documentclass[
  reprint, prx,
  superscriptaddress,
  amsmath,amssymb,
  aps,floatfix,
  showkeys,longbibliography]{revtex4-2}
\usepackage[dvipsnames]{xcolor}
\usepackage{tcolorbox}
\usepackage{wrapfig}
\usepackage{floatrow}
\usepackage[english]{babel}
\usepackage{graphicx}
\usepackage{nccmath}
\usepackage{float}
\usepackage{relsize}

\usepackage{cleveref}
\crefname{equation}{Eq.}{Eqs.}
\crefrangelabelformat{equation}{(#3#1#4--#5#2#6)}

\usepackage{amsmath}

\usepackage{mathrsfs}\newcommand{\rmd}{\mathrm{d}}
\usepackage{textcomp, gensymb}
\usepackage{nccmath}
\usepackage{mathtools} 
\newcommand{\lb}{\langle}
\newcommand{\rb}{\rangle}

\global\long\def\Cov{\mathrm{Cov}}
\global\long\def\CV{\mathrm{CV}}

\global\long\def\Var{\mathrm{Var}}

\usepackage{lipsum}

\begin{document}

\title{Using stationary information flows to prove kinetic uncertainty relations in biochemical control systems}

\author{Ryan Ripsman}
\altaffiliation[Present address: ]{Department of Cellular and Physiological Science,  University of British Columbia, Vancouver, BC V6T 1Z3, Canada}
\affiliation{Department of Physics, University of Toronto, Toronto, ON M5S 1A7, Canada}

\author{Brayden Kell}
\affiliation{Department of Physics, University of Toronto, Toronto, ON M5S 1A7, Canada}
\affiliation{Department of Chemical \& Physical Sciences, University of Toronto, Mississauga, ON L5L 1C6, Canada}

\author{Andreas Hilfinger}
\email{andreas.hilfinger@utoronto.ca}
\affiliation{Department of Physics, University of Toronto, Toronto, ON M5S 1A7, Canada}
\affiliation{Department of Chemical \& Physical Sciences, University of Toronto, Mississauga, ON L5L 1C6, Canada}

\begin{abstract}
Many cellular components are present in such low numbers that individual stochastic production and degradation events lead to significant fluctuations in molecular abundances. 
Although feedback control can, in principle, suppress such low-copy-number fluctuations, general rules have emerged that suggest fundamental performance constraints %
 on feedback control in biochemical %
systems.
In particular, previous work has conjectured that reducing abundance fluctuations in one component requires at least one sacrificial component with increased variability in arbitrary reaction networks of any size. Here, we present an exact and general proof of this statement based on probability current decompositions of mutual information rates between molecular abundances. This suggests that variability in cellular components is necessary for cellular control and that fluctuating components do not necessarily generate cellular ``noise'' but may correspond to control molecules that are involved in removing ``noise'' from other cellular components.
\end{abstract}


\maketitle

\textit{Introduction---}Biochemical processes involve low abundance species~\cite{Guptasarma1995,Bar-Even2006,Picotti2009} such that stochastic synthesis and degradation of individual molecules leads to significant abundance fluctuations or ``noise'' in biological cells~\cite{McAdams1999, Thattai2001, Elowitz2002, kaern2005, Bar-Even2006, Tsimring2014}. When transmitted through reaction networks, these low-copy-number %
effects lead to significant variability for downstream cellular components, regardless of their copy numbers~\cite{Paulsson2004}. In principle, negative feedback encoded through biochemical reactions can suppress such molecular fluctuations~\cite{Becskei2000}.
Yet many cellular components exhibit significant fluctuations~\cite{Ozbudak2002, Elowitz2002, Blake2003, Raser2004, Bar-Even2006}, %
which raises a fundamental question: If biological systems can in principle evolve arbitrarily precise control systems, why is there so much cellular ``noise''? One possible explanation is that basic constraints on the structure of biochemical control systems fundamentally limit their ability to suppress molecular fluctuations~\cite{Lestas2010, Hilfinger2016a, Kell2025}. 

Motivated by genetic feedback control, we consider
biochemical control loops that involve at least one intermediate  species. %
Such intermediate species are typically present even in ``minimal'' genetic control loops, such as transcriptional autorepression~\cite{Becskei2000}, where information about changes in protein abundance propagates through probabilistic gene activation, transcription and translation events~\cite{Lestas2010}. Similarly, ``autocatalytic'' processes such as the synthesis of ribosomes or RNA polymerases involve several intermediate steps in such a way that the rate of the final control step does not \emph{directly} depend on the abundance of the controlled species~\cite{Roy2021}. %

We present an analytical proof that regardless of control functions and topology, biochemical signalling cascades cannot simultaneously reduce fluctuations in all components, i.e., reducing fluctuations in one component requires at least one sacrificial component with increased fluctuations.

\begin{figure}[h]
    \centering
    \includegraphics[width=\linewidth]{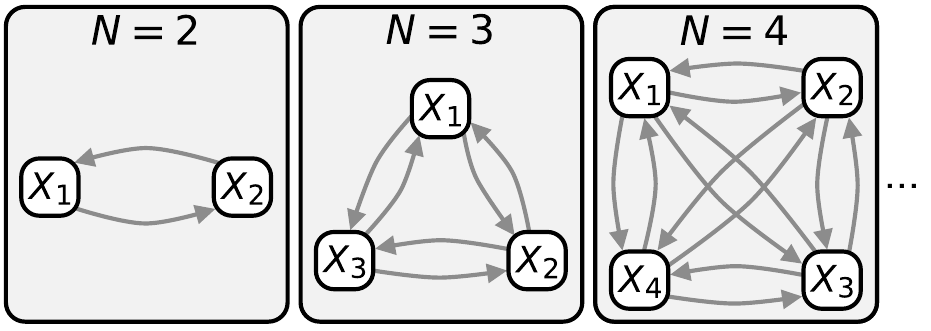}%
    \caption{\textbf{Example control networks within the general class of Eq.~\ref{eq: arbitrary network}.} We consider arbitrary $N$-dimensional control networks in which each component undergoes first order degradation and is allowed (but not required) to affect the production rate of any other component, see Eq.~\ref{eq: arbitrary network}. This allows for unspecified complexity in control topologies and control functions, but excludes direct feedback in which a biochemical species \emph{directly} affects its own production. Furthermore, we assume that each chemical reaction changes the molecular abundance of only one species.
    We illustrate maximally connected topologies involving $N=2,3,4$ components.}
    \label{fig: network diagram}
\end{figure}

\textit{Results---}We consider arbitrary $N$-dimensional reaction networks with molecular species $X_1,\ldots, X_N$ that influence each other's abundance dynamics through arbitrary production rates and undergo first order degradation (Fig.~\ref{fig: network diagram}). We place no constraints on the topology of the reaction network or on the form of rate functions, except that feedback must pass through intermediate components, i.e., the production rate of component $X_i$ is an arbitrary function 
of the state of all components in the network excluding $X_i$ itself. We also allow for arbitrary reaction timescales, which are determined by the average molecular lifetimes $\tau_i$, but we consider only networks in which each chemical reaction changes the abundance of only one species.

\newpage
Under the assumption that the system is well-mixed, molecules of each species $X_i$ are then produced and degraded according to the following probabilistic reaction events
\begin{equation}
\label{eq: arbitrary network}
    x_i \overset{f_i\left(\boldsymbol{x}^{(i)}\right)}{\xrightarrow{\hspace*{1.1cm}}} x_i + 1 \ , \qquad x_i \overset{x_i/\tau_i}{\xrightarrow{\hspace*{1.1cm}}} x_i - 1 \ , \vspace{0.5mm} 
\end{equation}
where $\boldsymbol{x}^{(i)} \coloneqq (x_1,\ldots, x_{i-1},x_{i+1},\ldots,x_N)$ denotes the vector of molecular abundances excluding $X_i$,
and the reaction rates $f_i\left(\boldsymbol{x}^{(i)}\right)$ and $x_i/\tau_i$ determine the continuous-time transition probabilities for the discrete production and degradation events.

We consider the stationary abundance fluctuations of all stable stochastic processes described by the reactions of Eq.~\ref{eq: arbitrary network}. The stationary state corresponds to the long-time behavior of stable stochastic systems, where the joint probability distribution of states no longer changes in time. 
For each component $X_i$ in a given control network, we quantify the abundance fluctuations through the coefficient of variation \mbox{$\CV_i := \sqrt{\Var(x_i)}/\lb x_i \rb$}, where the variance and mean are taken with respect to the stationary distribution. We compare the fluctuations of each $X_i$ within the control network to the Poisson fluctuations $\CV_i=\sqrt{1/\langle x_i \rangle}$ it would exhibit at the same mean abundance in the absence of control.

Previous work~\cite{yan2019kinetic} conjectured that it is impossible to simultaneously suppress noise in all components below their uncontrolled levels for the broad class of stochastic reaction networks described by Eq.~\ref{eq: arbitrary network} -- i.e., if there exists a component $X_i$ with $\CV_i < \sqrt{1/\langle x_j \rangle}$ then there must exist at least one component $X_j$ with
\begin{equation}
\label{eq: conjecture}
 \CV_j > \sqrt{1/\langle x_j \rangle}\ ,
\end{equation}
regardless of the complexity of the network, choice of control functions, or reaction timescales. 

This conjecture is supported analytically for systems of arbitrary size in the asymptotic limit of high abundances~\cite{vank1992, Paulsson2004}, %
and numerically for small systems through a large number of simulations %
~\cite{yan2019kinetic}. However, real biochemical systems often have low abundances and numerical support becomes computationally intractable for large systems with high-dimensional parameter spaces. Even for small $N$, there is no guarantee that the observed limits for a large set of simulated control systems reflect fundamental performance limits of control %
rather than limits of our imagination.

Here, we analytically prove the conjecture of Eq.~\ref{eq: conjecture}, i.e., we show without approximations that noise suppression in arbitrarily complex control networks described by the reactions of Eq.~\ref{eq: arbitrary network} requires at least one ``sacrificial'' component with increased fluctuations compared to their uncontrolled fluctuations. Our approach is based on a decomposition of a mutual information rate%
, closely following previous work on thermodynamic systems with %
multipartite state spaces~\cite{Horowitz2014}.

\begin{figure*}
    \centering
    \includegraphics[width=0.85\linewidth]{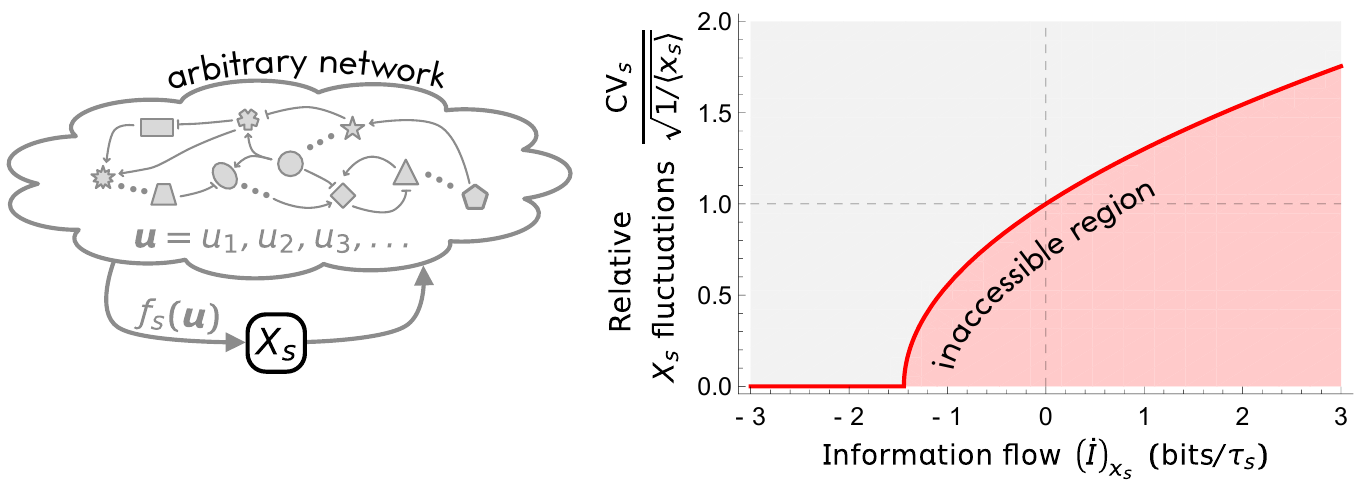}
    \vspace{-.5em}
    \caption{\textbf{Stationary information flow limits stationary abundance fluctuations in arbitrary reaction networks.} We consider any component $X_s$ undergoing first order degradation and subject to an arbitrarily complicated control network with unspecified dynamics, see Eq.~\ref{eq: more general class}. The unspecified control network sets the production rate $f_s(\boldsymbol{u})$ of $X_s$ molecules and the abundance of $X_s$ molecules can influence the control network in arbitrary ways. %
    The inequality of Eq.~\ref{eq: CV-MI flow bound} bounds the relative noise control in terms of the stationary mutual information flow between $X_s$ and the unspecified network resulting from jumps in the $X_s$ abundance. Information flow is plotted in units of bits per 
    average lifetime of $X_s$ molecules, $\tau_s$. While the plotted bound is analytically proven and cannot be overcome by any control system, it is not necessarily tight and more stringent bounds might apply. The 
    cusp at 
    $(\dot{I})_{x_s} \approx -1.4 \ \mathrm{bits}/\tau_s$
    should thus not be interpreted as a critical point below which noise suppression is unconstrained.} %
    \label{fig: CV-MI flow bound}
\end{figure*}

We first establish a lemma that constrains the noise suppression of any component $X_s$ undergoing first order degradation and whose production rate depends on an arbitrary network
\begin{equation}\label{eq: more general class}
    x_s \overset{f_s\left(\boldsymbol{u}\right)}{\xrightarrow{\hspace*{1.1cm}}} x_s + 1 \ , \qquad x_s \overset{x_s/\tau_s}{\xrightarrow{\hspace*{1.1cm}}} x_s - 1 ,
\end{equation}
where $\boldsymbol{u}$ is the dynamical state of the unspecified network $\boldsymbol{U}$, see Fig.~\ref{fig: CV-MI flow bound}. Note, this class of systems contains the systems of Eq.~\ref{eq: arbitrary network}, but is even more general since the dynamics of $\boldsymbol{u}$ are left completely unspecified. 

For all such systems, the stationary state fluctuations of component $X_s$ satisfy
\begin{equation}\label{eq: X_s fluctuation balance}
    \left(\frac{\CV_s}{\sqrt{1/\lb x_s \rb}}\right)^2 = 1 + \frac{\Cov\left(f_s(\boldsymbol{u}),x_s \right)}{\lb f_s(\boldsymbol{u}) \rb}\,
\end{equation}
regardless of the details of the unspecified control network~\cite{Hilfinger2016a}. Noise suppression in $X_s$ thus corresponds to control systems that achieve a negative covariance term in Eq.~\ref{eq: X_s fluctuation balance}, which is bounded by stationary information flows as detailed next.

$\mathcal{X}$
The mutual information between the abundance of the specified component $X_s$ and the state of the unspecified network $\boldsymbol{U}$ is defined as
\begin{equation}\label{eq: MI}
    I(x_s; \boldsymbol{u}) = \sum_{x_s}\sum_{\boldsymbol{u}}P(x_s,\boldsymbol{u})\ln \frac{P(x_s,\boldsymbol{u})}{P(x_s)P(\boldsymbol{u})}\ .
\end{equation}
By assumption, the reactions of Eq.~\ref{eq: more general class} denote all reactions that change the abundance of $X_s$, and none of those reactions directly change abundances in the unspecified network $\boldsymbol{u}$. Thus  
the mutual information rate can be decomposed into information flows quantifying how jumps in the subsystems $x_s$ and $\boldsymbol{u}$ %
affect the mutual information between them~\cite{Horowitz2014, Horowitz2015}, i.e., $\rmd_t I(x_s; \boldsymbol{u}) = (\dot{I})_{x_s} + (\dot{I})_{\boldsymbol{u}}$.

In particular, the flow due to $x_s$ jumps (see Appendix A) is given by
\begin{multline}\label{eq: x_s info flow MT}
    \left(\dot{I}\right)_{x_s} \coloneqq \sum_{x_s}\sum_{\boldsymbol{u}} P(x_s,\boldsymbol{u})f_s(\boldsymbol{u})\ln \frac{P(\boldsymbol{u}|x_s + 1)}{P(\boldsymbol{u}|x_s)} \\ 
    + \sum_{x_s}\sum_{\boldsymbol{u}} P(x_s,\boldsymbol{u})\frac{x_s}{\tau_s}\ln \frac{P(\boldsymbol{u}|x_s - 1)}{P(\boldsymbol{u}|x_s)}\ ,
\end{multline}
where $P(x_s,\boldsymbol{u})$ is the joint distribution of states for the combined system including both the specified component $X_s$ and the unspecified network $\boldsymbol{U}$, and $P(\boldsymbol{u}|x_s)$ is the conditional probability of observing the unspecified network in state $\boldsymbol{u}$ given the abundance of $X_s$ is $x_s$. 

We next show that the size of spontaneous $X_s$ abundance fluctuations in the stationary state is limited by the information flow $(\dot{I})_{x_s}$.
We achieve this by applying the log sum inequality~\cite{Cover2001} along with basic inequalities for logarithms of non-negative numbers and stationary state probability flux balance relations (see Appendix B) to Eq.~\ref{eq: x_s info flow MT} and obtain
\begin{equation}\label{eq: X_s MI flow bound}
    (\dot{I})_{x_s} \leq \frac{1}{\tau_s}\frac{\Cov\left(f_s(\boldsymbol{u}),x_s \right)}{\lb f_s(\boldsymbol{u}) \rb}\ .
\end{equation}
Combining Eqs.~\ref{eq: X_s fluctuation balance} and \ref{eq: X_s MI flow bound} then establishes the following bound on the noise control in $X_s$
\begin{equation}\label{eq: CV-MI flow bound}
    \left(\frac{\CV_s}{\sqrt{1/\lb x_s \rb}}\right)^2 \geq 1 + \tau_s (\dot{I})_{x_s}\ .
\end{equation}

The left-side of Eq.~\ref{eq: CV-MI flow bound} quantifies the fluctuations of $X_s$ relative to the fluctuations of an uncontrolled birth-death process with the same mean abundance. %
Reducing $X_s$ fluctuations %
thus requires $(\dot{I})_{x_s} < 0$ (Fig.~\ref{fig: CV-MI flow bound}). This means that jumps in the abundance of $X_s$ must, on average, decrease the mutual information $I(x_s;\boldsymbol{u})$~\cite{Horowitz2014}. %
This in turn implies $ (\dot{I})_{\boldsymbol{u}} >0$, since at stationarity $\mathrm{d}_t I(x_s; \boldsymbol{u}) = 0$ . 
Jumps in the state of the unspecified network must thus, on average, increase the mutual information $I(x_s;\boldsymbol{u})$ in order to reduce $X_s$ fluctuations (see Appendix A). In this way, the unspecified network can be intuitively interpreted as a control system whose state transitions tend to generate information about $X_s$~\cite{Horowitz2014}. A technically different but conceptually similar bound has been discussed in the context of buffering environmental fluctuations in biochemical sensory adaptation~\cite{Ito2015}. %

The lemma of Eq.~\ref{eq: CV-MI flow bound} applies to any component $X_i$ in systems described by Eq.~\ref{eq: arbitrary network}. %
Thus, to establish the conjecture of Eq.~\ref{eq: conjecture}, it %
suffices to show that noise suppression in one component requires that another component's state transitions must (on average) generate information about the rest of the network, i.e., that $(\dot{I})_{x_i} < 0$ for some component $X_i$ implies $(\dot{I})_{x_j} > 0$ for at least one other component $X_j$. To show this we consider the joint entropy 
\begin{equation}\label{eq: joint entropy defn}
    H(\boldsymbol{x}) = \sum_{\boldsymbol{x}} P(\boldsymbol{x}) \ln P(\boldsymbol{x})\ .
\end{equation}
We then show the following information flow balance holds for stationary systems where $\mathrm{d}_t H(\boldsymbol{x}) = 0$ (see Appendix C):
\begin{equation}\label{eq: info flow balance}
    \sum_{i=1}^N (\dot{I})_{x_i} = 0 \ ,
\end{equation}
where $(\dot{I})_{x_i}$ is the contribution of $x_i$ jumps to $\rmd_t I(x_i;\boldsymbol{x}^{(i)})$ and is defined analogously to $(\dot{I})_{x_s}$ (Eq.~\ref{eq: x_s info flow MT}). Combining Eqs.~\ref{eq: CV-MI flow bound} and \ref{eq: info flow balance}
then implies
\begin{equation}\label{eq: CV-tau bound}
    \mathlarger{\sum_{i=1}^N}\hspace{1mm}\frac{1}{\tau_i} \left[\left(\frac{\CV_i}{\sqrt{1/\lb x_i \rb}}\right)^2 - 1\right] \geq 0 \ .
\end{equation}

\begin{figure}[h]
    \centering
    \includegraphics[width=0.85\linewidth]{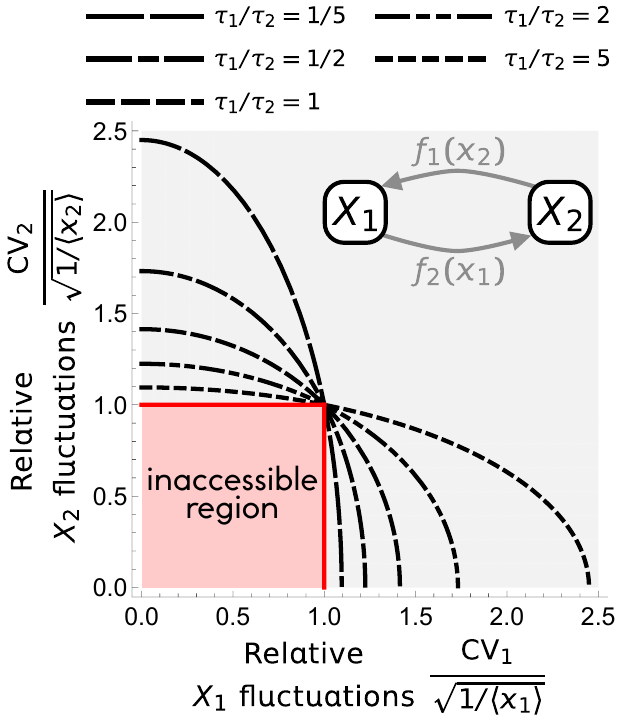}
    \caption{\textbf{Kinetic uncertainty relation for mutual control systems.} Illustrated are the analytically proven general bounds using an arbitrary two-component feedback loop where each component undergoes first order degradation and sets the production rate of the other component, i.e., the class of systems given by Eq.~\ref{eq: arbitrary network} with $N=2$. Eq.~\ref{eq: CV-tau bound} determines the space of allowed relative fluctuation values for a given ratio of the average molecular lifetimes $\tau_1/\tau_2$. Here we plot the lower bound on $X_2$ fluctuations for a given value of the $X_1$ fluctuations for a range of timescale ratios. The region outlined by the solid red line is excluded for all timescale ratios and is approached only in the limit of infinite timescale separation. In the general $N$-component case, this inaccessible region becomes a hypercube, such that at least one component must have increased fluctuations, see Eq.~\ref{eq: CV-tau bound}.} %
    \label{fig: timescales}
\end{figure}

The class of reaction networks described by Eq.~\ref{eq: arbitrary network} thus does not admit stationary states where $\CV_i < \sqrt{1/\lb x_i \rb}$ for every component $X_i$. This completes the proof of Eq.~\ref{eq: conjecture}: noise suppression requires at least one sacrificial component with increased fluctuations for any network described in the class of systems specified by Eq.~\ref{eq: arbitrary network}. 

The bound of Eq.~\ref{eq: CV-tau bound} additionally establishes that the noise penalty required for noise suppression can only vanish in the limit of infinite timescale separation. More precisely, the variability of the noisiest component(s) can approach Poisson fluctuations only in the limit that at least one of those components exhibits infinitely fast dynamics relative to every component with sub-Poisson fluctuations. Fig.~\ref{fig: timescales} illustrates this point in the example of two-component control networks. For finite timescale ratios $\tau_i/\tau_j$, previous work conjectured more severe bounds than Eq.~\ref{eq: CV-tau bound} and %
predicted that eliminating fluctuations entirely in one component requires diverging fluctuations in another~\cite{yan2019kinetic}. Given numerical support of such conjectured bounds for small systems~\cite{yan2019kinetic}, it thus appears that while Eq.~\ref{eq: CV-tau bound} rigorously establishes fundamental bounds, the proven bounds are loose for finite timescale separation. %

\textit{Discussion---}Despite their complexity and often unknown parameters, noise properties of intracellular reaction networks may be explained by general
performance limits~\cite{Lestas2010, Hilfinger2016a, Kell2025, aatay2009}%
. For example, previous work has shown that information loss along signalling steps with finite rates places severe limits on the suppression of molecular fluctuations~\cite{Lestas2010}. In contrast, developments in stochastic thermodynamics have established bounds on fluctuations in terms of dissipation in non-equilibrium reaction networks~\cite{Barato2015, Gingrich2016, Barato2017,Gingrich2017,Chun2023}. 

The bounds derived here are based solely on the structure of the reaction network, and cannot be overcome by any degree of non-equilibrium driving. Much stricter bounds apply to systems at thermodynamic equilibrium. Specifically, the fluctuations of systems that satisfy detailed balance and are described by Eq.~\ref{eq: arbitrary network}, must exceed Poisson levels for \emph{all} components, since noise suppression requires a non-zero probability current. Approaching the fundamental limits derived here thus 
requires control systems out of thermodynamic equilibrium. 

The information flow decompositions used here appear in stochastic thermodynamics, where they bound the dissipation of interacting non-equilibrium systems~\cite{Hartich2014, Horowitz2014a, Horowitz2014, Horowitz2015, Shiraishi2015, Parrondo2015, Penocchio2022, Ehrich2023, Grelier2024, Leighton2024, Leighton2025, Buisson2025}. It is therefore tempting to exploit this connection~\cite{Horowitz2014, Horowitz2015} to cast Eq.~\ref{eq: CV-MI flow bound} as a lower bound on fluctuations for a given level of dissipation. %
This, however, would be misleading, as we do not model elementary reactions with forward and reverse rates based on chemical potentials~\cite{Mou1986, Schmiedl2007}. Rather, as is often the case in biological systems, the degradation reaction here does not correspond to %
synthesis in reverse, but can rather be a driven process in its own right~\cite{yan2019kinetic}. In this way, the elementary reactions implicitly underlying the mathematically coarse-grained transitions considered here can be driven arbitrarily far from thermodynamic equilibrium. Deriving stricter bounds on fluctuations for a given level of dissipation in molecular subsystems would benefit from an information-thermodynamic framework for non-elementary \emph{stochastic} reaction networks~\cite{Avanzini2020, Penocchio2022}. 

Our results place hard bounds on fluctuation control within a very general class of stochastic reaction networks. %
Nevertheless, our results permit that $N$-component biochemical control systems can, in principle, achieve sub-Poisson fluctuations for $N-1$ components as long as \emph{one} component has at least Poisson fluctuations. This statement does not necessarily explain the myriad of noisy components observed in cells. We are, however, very general in our assumptions in allowing for every component to influence every other component's synthesis rate arbitrarily with arbitrary reaction timescales. Previous work suggests much stricter fluctuation bounds when the topology of the network is constrained~\cite{yan2019kinetic}. For example, a large number of numerical simulations for small systems support the conjecture that \emph{at most} one component can exhibit sub-Poisson fluctuations in feedback rings~\cite{yan2019kinetic}. Future work is needed to extend our analytical tools to incorporate constraints on the topology of the reaction network and to 
prove this 
 conjecture.%
 
While we consider a mathematically very broad class of systems%
, there are important types of biochemical reaction networks which are excluded from Eq.~\ref{eq: arbitrary network}. Specifically, our analysis excludes systems in which the abundance of multiple components changes in the same reaction. Examples of such biochemical processes include conversion due to conformational switching or assembly and disassembly of multi-subunit complexes. Future work is %
required to analyze if and how the kinetic uncertainty relations discussed here apply to systems which include such biochemical motifs. Given that efficient complex-formation requires diverging subunit fluctuations in the absence of feedback~\cite{Hilfinger2016a, Kell2025}, such extensions may explain conjectured noise bounds for other components within classes of feedback loops embedding complex-formation motifs~\cite{Kell2023}.

\textit{Acknowledgements---}We thank Seshu Iyengar for drawing our attention to several key papers that motivated and informed the development of this work. We also thank Glenn Vinnicombe for a key suggestion to streamline the presented analytical proof.
The work was supported by Natural Sciences and Engineering Research Council of Canada (NSERC) Discovery Grants RGPIN-2025-06695 and RGPIN-2019-06443. B.K.~gratefully acknowledges funding from a University of Toronto Faculty of Arts \& Science Top Doctoral Fellowship and an Ontario Graduate Scholarship. %

\textit{Author Contributions---} R.R.~-- Formal analysis (mathematical derivations), methodology, writing -- review \& editing. B.K.~-- Writing -- original draft, writing -- review \& editing, visualization, formal analysis (mathematical derivations). A.H.~-- Conceptualization, funding acquisition, supervision, writing -- original draft, writing -- review \& editing. 

\appendix 

\section*{APPENDIX}

\textit{Appendix A: Derivation of Eq.~\ref{eq: x_s info flow MT}---}Regardless of the dynamics of $\boldsymbol{u}$, if the dynamics of $X_s$ follow Eq.~\ref{eq: more general class}, then the time-dependent joint probability distribution $P(x_s,\boldsymbol{u})$ evolves according to a master equation~\cite{vank1992} of the form
\begin{equation}\label{eq: x_s ME}
    \frac{\rmd P(x_s,\boldsymbol{u})}{\rmd t} = \sum_{x_s'}J_{x_s,x_s'}^{\boldsymbol{u}} + \sum_{\boldsymbol{u}'}J_{\boldsymbol{u},\boldsymbol{u}'}^{x_s}%
\end{equation}
where the probability current into the combined state $(x_s,\boldsymbol{u})$ along the jump $x_s' \to x_s$ is
\begin{multline}\label{eq: x_s current}
    J_{x_s,x_s'}^{\boldsymbol{u}} \\ 
    \coloneqq \begin{cases} 
    \begin{aligned}
        f_s(\boldsymbol{u})P(x_s - 1, \boldsymbol{u}) &- \frac{x_s}{\tau_s}P(x_s, \boldsymbol{u})\ ,\\[3pt] & \ \ \quad \mathrm{if \ } x_s' = x_s-1  
    \end{aligned}\\
    \begin{aligned}
        \frac{x_s + 1}{\tau_s}P(x_s+1, \boldsymbol{u})  &- f_s(\boldsymbol{u})P(x_s,\boldsymbol{u})\ ,\\  &  \hspace{0.3mm} \quad \mathrm{if \ } x_s' = x_s+1 
    \end{aligned}\\
        0 \ , \quad \hspace{25.3 mm} \mathrm{else}
    \end{cases}
\end{multline}
The probability current $J_{\boldsymbol{u},\boldsymbol{u}'}^{x_s}$ due to jumps in the unspecified network $\boldsymbol{U}$ is arbitrary. 

Applying standard differentiation procedures and the law of total probability shows that \mbox{$\sum_{x_s}\sum_{\boldsymbol{u}} P(x_s,\boldsymbol{u})\rmd_t\ln \left( P(x_s,\boldsymbol{u})/P(x_s)P(\boldsymbol{u})\right) = 0$}. Thus the time derivative of the mutual information $I(x_s;\boldsymbol{u})$, as defined in Eq.~\ref{eq: MI}, is
\begin{multline}\label{eq: X_s info flow decomp}
    \frac{\mathrm{d}I(x_s; \boldsymbol{u})}{\mathrm{d}t} = \underbrace{\sum_{x_s}\sum_{\boldsymbol{u}}\sum_{x_s'}J_{x_s,x_s'}^{\boldsymbol{u}} \ln \frac{P(x_s,\boldsymbol{u})}{P(x_s)P(\boldsymbol{u})}}_{\eqcolon \left(\dot{I}\right)_{x_s}}\\ 
    + \underbrace{\sum_{x_s}\sum_{\boldsymbol{u}}\sum_{\boldsymbol{u}'}J_{\boldsymbol{u},\boldsymbol{u}'}^{x_s} \ln \frac{P(x_s,\boldsymbol{u})}{P(x_s)P(\boldsymbol{u})}}_{\eqcolon \left(\dot{I}\right)_{\boldsymbol{u}}} \ .
\end{multline}

Since $x_s$ takes jumps with unit step sizes and currents satisfy \mbox{$J_{x_s,x_s+1}^{\boldsymbol{u}} = - J_{x_s+1,x_s}^{\boldsymbol{u}}$}, the information flow due to jumps in $x_s$ can be written
\begin{equation}\label{eq: info flow defn}
    \left(\dot{I}\right)_{x_s} = \sum_{x_s}\sum_{\boldsymbol{u}} J_{x_s+1,x_s}^{\boldsymbol{u}}\ln \frac{P(\boldsymbol{u}|x_s + 1)}{P(\boldsymbol{u}|x_s)} \ .
\end{equation}
Substituting the probability current Eq.~\ref{eq: x_s current} then gives the following decomposition of the information flow:
\begin{multline}\label{eq: X_s info flow decomposition}
    \left(\dot{I}\right)_{x_s} = \underbrace{\sum_{x_s}\sum_{\boldsymbol{u}} P(x_s,\boldsymbol{u})f_s(\boldsymbol{u})\ln \frac{P(\boldsymbol{u}|x_s + 1)}{P(\boldsymbol{u}|x_s)}}_{\eqcolon \left(\dot{I}\right)_{x_s}^{+}} \\ 
    + \underbrace{\sum_{x_s}\sum_{\boldsymbol{u}} P(x_s,\boldsymbol{u})\frac{x_s}{\tau_s}\ln \frac{P(\boldsymbol{u}|x_s - 1)}{P(\boldsymbol{u}|x_s)}}_{\eqcolon \left(\dot{I}\right)_{x_s}^{-}}\ ,
\end{multline}
where $ (\dot{I})_{x_s}^\pm$ correspond to the contributions from $X_s$ production ($+$) and degradation ($-$) events in Eq.~\ref{eq: more general class}.

\textit{Appendix B: Derivation of Eq.~\ref{eq: X_s MI flow bound}---}
By the definition of conditional probabilities, we have
\begin{multline}\label{eq: I-dot -}
        \left(\dot{I}\right)_{x_s}^{-} \\ = -\sum_{x_s}P(x_s)\frac{x_s}{\tau_s}\sum_{\boldsymbol{u}}P(\boldsymbol{u}|x_s)\ln \frac{P(\boldsymbol{u}|x_s)}{P(\boldsymbol{u}|x_s - 1)} \ ,
\end{multline}
where $\left(\dot{I}\right)_{x_s}^{-}$ is defined in Eq.~\ref{eq: X_s info flow decomposition}. 
Additionally, we have
\begin{multline}
    \sum_{\boldsymbol{u}}P(\boldsymbol{u}|x_s)\ln \frac{P(\boldsymbol{u}|x_s)}{P(\boldsymbol{u}|x_s - 1)} \\ \geq \underbrace{\sum_{\boldsymbol{u}}P(\boldsymbol{u}|x_s)}_{=1}\underbrace{\ln \frac{\sum_{\boldsymbol{u}} P(\boldsymbol{u}|x_s)}{\sum_{\boldsymbol{u}} P(\boldsymbol{u}|x_s - 1)}}_{=0} = 0 \ , 
\end{multline}
where the inequality is the log sum inequality~\cite{Cover2001}. We thus have 
\begin{equation}\label{eq: I- bound}
    \left(\dot{I}\right)_{x_s}^{-} \leq 0\ .
\end{equation}

A similar application of the log sum inequality to the definition of $\left(\dot{I}\right)_{x_s}^{+}$ gives
\begin{equation}\label{eq: log sum}
    \left(\dot{I}\right)_{x_s}^{+} \leq \sum_{x_s}P(x_s) \lb f_s(\boldsymbol{u}) | x_s \rb \ln \frac{\lb f_s(\boldsymbol{u}) | x_s + 1 \rb}{\lb f_s(\boldsymbol{u}) | x_s \rb}\ .
\end{equation}
The inequalities $1 - 1/y \leq \ln y \leq y - 1$ hold for any $y \geq 0$. Rescaling the numerator and denominator of the logarithm argument above by $\lb f_s(\boldsymbol{u})\rb$ and applying these standard inequalities leads to 
\begin{multline}\label{eq: I+ bound}
    \left(\dot{I}\right)_{x_s}^{+} \leq \frac{1}{\lb f_s(\boldsymbol{u})\rb}\sum_{x_s}P(x_s)\lb f_s(\boldsymbol{u})|x_s\rb \lb f_s(\boldsymbol{u})|x_s+1\rb\\
    - \lb f_s(\boldsymbol{u})\rb
     \ . 
\end{multline}
Summing the master equation (see Eqs.~\ref{eq: x_s ME} and \ref{eq: x_s current}) over $\boldsymbol{u}$ gives the following probability flux balance at stationarity:
\begin{multline}
    \lb f_s(\boldsymbol{u}) |x_s-1 \rb P(x_s-1) + \frac{x_s + 1}{\tau_s}P(x_s + 1) \\ 
    = \left(\lb f_s(\boldsymbol{u}) | x_s \rb + \frac{x_s}{\tau_s} \right)P(x_s)\ .
\end{multline}
It can then be shown inductively (or verified by substitution) that the stationary marginal distribution $P(x_s)$ satisfies
\begin{equation}\label{eq: balance relation}
    \langle f_s(\boldsymbol{u})|x_s \rangle P(x_s) = \frac{x_s+1}{\tau_s}P(x_s+1)\ .
\end{equation}
Thus,
\begin{equation}
\begin{aligned}
    \sum_{x_s}P(x_s)&\lb f_s(\boldsymbol{u})|x_s\rb \lb f_s(\boldsymbol{u})|x_s+1\rb 
    \\
    &= \sum_{x_s}\frac{x_s+1}{\tau_s}P(x_s+1) \lb f_s(\boldsymbol{u})|x_s+1\rb \\
    &= \frac{\lb x_s \lb f_s(\boldsymbol{u})|x_s\rb\rb}{\tau_s} = \frac{\lb x_s f_s(\boldsymbol{u})\rb}{\tau_s}\ . %
\end{aligned}
\end{equation}
Combined with Eq.~\ref{eq: I+ bound}, we have
\begin{equation}\label{eq: I+ bound final}
    \left(\dot{I}\right)_{x_s}^{+} \leq \underbrace{\frac{\lb x_s f_s(\boldsymbol{u})\rb}{\tau_s\lb f_s(\boldsymbol{u})\rb} - \lb f_s(\boldsymbol{u})\rb}_{= \Cov(f_s(\boldsymbol{u}),x_s)/\lb f_s(\boldsymbol{u}) \rb \tau_s}\ .
\end{equation}
where the equality below the brace follows because stationary flux balance requires that \mbox{$\lb f_s(\boldsymbol{u}) \rb = \lb x_s \rb/\tau_s$} (e.g. see~\cite{Hilfinger2016a}) and, by definition, \mbox{$\Cov(f_s(\boldsymbol{u}),x_s) = \lb x_s f_s(\boldsymbol{u})\rb - \lb f_s(\boldsymbol{u})\rb \lb x_s \rb$}. Combining Eqs.~\ref{eq: I- bound} and \ref{eq: I+ bound final} with the decomposition Eq.~\ref{eq: X_s info flow decomposition} then implies Eq.~\ref{eq: X_s MI flow bound}, as required.

\textit{Appendix C: Derivation of Eq.~\ref{eq: info flow balance}---}The time-dependent joint probability distribution $P(\boldsymbol{x})$ for the class of Markov processes described by Eq.~\ref{eq: arbitrary network} evolves according to a master equation~\cite{vank1992} of the form
\begin{equation}\label{eq: ME}
    \frac{\rmd P(\boldsymbol{x})}{\rmd t} = \sum_{i = 1}^N \sum_{x_i'} J_{x_i,x_i'}^{\boldsymbol{x}^{(i)}} \ , 
\end{equation}
where

\begin{equation}\label{eq: currents}
    J_{x_i,x_i'}^{\boldsymbol{x}^{(i)}}
    \coloneqq \begin{cases} 
    \begin{aligned}
        f_i(\boldsymbol{x}^{(i)})P(\boldsymbol{x}-\boldsymbol{e}_i) &- \frac{x_i}{\tau_i}P(\boldsymbol{x})\ ,\\[3pt] & \ \mathrm{if \ } x_i' = x_i-1  
    \end{aligned}\\
    \begin{aligned}
        \frac{x_i + 1}{\tau_i}P(\boldsymbol{x}+\boldsymbol{e}_i)  &- f_i(\boldsymbol{x}^{(i)})P(\boldsymbol{x}) \ ,\\  &  \hspace{2.4  mm}  \mathrm{if \ } x_i' = x_i+1 
    \end{aligned}\\
        0 \ , \quad \hspace{19.8 mm} \mathrm{else}
    \end{cases}
\end{equation}
is the probability current into state $\boldsymbol{x}$ along the jump $x_i' \to x_i$ and $\boldsymbol{e}_i \in \mathbb{Z}^N$ is the standard unit basis vector in the direction of increasing $x_i$. 

Following an analogous argument to that which led to Eqs.~\ref{eq: X_s info flow decomp} and \ref{eq: info flow defn}, we have
\begin{equation}
    \frac{\mathrm{d}H(\boldsymbol{x})}{\mathrm{d}t} = \sum_{i = 1}^N \sum_{\boldsymbol{x}} J_{x_i+1,x_i}^{\boldsymbol{x}^{(i)}} \ln \frac{P(\boldsymbol{x} + \boldsymbol{e}_i)}{P(\boldsymbol{x})}\ ,
\end{equation}
where $H(\boldsymbol{x})$ is the joint entropy defined in Eq.~\ref{eq: joint entropy defn}. From the definition of conditional probabilities it then follows that
\begin{multline}
    \frac{\mathrm{d}H(\boldsymbol{x})}{\mathrm{d}t} = \sum_{i=1}^N \left(\dot{I}\right)_{x_i} \\ - \sum_{i=1}^N \sum_{\boldsymbol{x}} J_{x_i+1,x_i}^{\boldsymbol{x}^{(i)}} \ln \frac{P(x_i)}{P(x_i+1)}\ ,
\end{multline}
where 
\begin{equation}\label{eq: X_i info flow defn}
    \left(\dot{I}\right)_{x_i} = \sum_{x_i}\sum_{\boldsymbol{x}^{(i)}} J_{x_i+1,x_i}^{\boldsymbol{x}^{(i)}}\ln \frac{P(\boldsymbol{x}^{(i)}|x_i + 1)}{P(\boldsymbol{x}^{(i)}|x_i)} \ .
\end{equation}
Then completing the sum over $\boldsymbol{x}^{(i)}$ gives
\begin{multline}\label{eq: dH/dt}
    \frac{\mathrm{d}H(\boldsymbol{x})}{\mathrm{d}t} = \sum_{i=1}^N \left(\dot{I}\right)_{x_i} \\ - \sum_{i=1}^N \sum_{x_i} \Tilde{J}_{x_i+1,x_i} \ln \frac{P(x_i)}{P(x_i+1)}\ ,
\end{multline}
where
\begin{equation}
\begin{aligned}
    \Tilde{J}_{x_i+1,x_i} &\coloneqq \sum_{\boldsymbol{x}^{(i)}} J_{x_i+1,x_i}^{\boldsymbol{x}^{(i)}} \\
    &= \langle f_i(\boldsymbol{x}^{(i)})|x_i \rangle P(x_i) - \frac{x_i+1}{\tau_i}P(x_i+1)\ .
\end{aligned}
\end{equation}
The same steps leading to the conditional probability flux balance of Eq.~\ref{eq: balance relation} apply to the conditionally averaged dynamics of any $X_i$ in the class of systems given by Eq.~\ref{eq: arbitrary network}. Therefore, at stationarity we have $\Tilde{J}_{x_i+1,x_i} = 0$. Eq.~\ref{eq: info flow balance} then follows  from Eq.~\ref{eq: dH/dt} at stationarity, since $\rmd_t H(\boldsymbol{x}) = 0$ for stationary systems. 

\bibliography{bibliography}

\end{document}